%% file: main.tex
\documentclass{article}
\usepackage{ijcai23}

\usepackage{times}
\usepackage{soul}
\usepackage{url}
\usepackage[hidelinks]{hyperref}
\usepackage[utf8]{inputenc}
\usepackage[small]{caption}
\usepackage{graphicx}
\usepackage{color}
\usepackage{amsmath}
\usepackage{amsthm}
\usepackage{booktabs}
\usepackage{algorithm}
\usepackage{algorithmic}

\usepackage{pifont}
\usepackage{booktabs}
\usepackage{array} 
\usepackage{multirow}
\usepackage{caption}
\usepackage[switch]{lineno}
\usepackage{lipsum}
\usepackage[most]{tcolorbox}

\definecolor{myblue}{RGB}{0,92,169}

\newtcolorbox{principle-1}{
  enhanced,
  colback=myblue!10,
  colframe=myblue!30,
  fonttitle=\bfseries,
  title=Principle 1,
  arc=0pt,
  outer arc=0pt,
  boxrule=1pt
}

\newtcolorbox{principle-2}{
  enhanced,
  colback=myblue!10,
  colframe=myblue!30,
  fonttitle=\bfseries,
  title=Principle 2,
  arc=0pt,
  outer arc=0pt,
  boxrule=1pt
}

\newtcolorbox{principle-3}{
  enhanced,
  colback=myblue!10,
  colframe=myblue!30,
  fonttitle=\bfseries,
  title=Principle 3,
  arc=0pt,
  outer arc=0pt,
  boxrule=1pt
}

\definecolor{mygreen}{RGB}{102,204,0}

\newtcolorbox{limitationbox}{
  enhanced,
  colback=white,
  colframe=mygreen!70,
  coltitle=mygreen!20!black,
  fonttitle=\bfseries,
  title=Limitation,
  arc=0pt,
  outer arc=0pt,
  boxrule=1pt,
  drop fuzzy shadow
}

\title{Breaking On-device Training Memory Wall: A Systematic Survey}

\author{
Shitian Li
\and
Chunlin Tian \thanks{Corresponding Author} \and
Kahou Tam\and
Rui Ma \And
Li Li \footnotemark[1]
\affiliations
University of Macau
\emails
\{dc12689\}@um.edu.mo, \{tianclin0212, wo133565\}@gmail.com,   \{yc27430, LLiLi\}@um.edu.mo
}

\begin{document}

\maketitle

\begin{abstract}
    On-device training has become an increasingly popular approach to machine learning, enabling models to be trained directly on mobile and edge devices. However, a major challenge in this area is the limited memory available on these devices, which can severely restrict the size and complexity of the models that can be trained. In this systematic survey, we aim to explore the current state-of-the-art techniques for breaking on-device training memory walls, focusing on methods that can enable larger and more complex models to be trained on resource-constrained devices. 
    Specifically, we first analyze the key factors that contribute to the phenomenon of memory walls encountered during on-device training. Then, we present a comprehensive literature review of on-device training, which addresses the issue of memory limitations. Finally, we summarize on-device training and highlight the open problems for future research.
    By providing a comprehensive overview of these techniques and their effectiveness in breaking memory walls, we hope to help researchers and practitioners in this field navigate the rapidly evolving landscape of on-device training.

\end{abstract}

\section{Introduction}

\input{Introduction}


\section{Memory Wall}
\input{MemoryWall}

\section{Existing Approaches Analysis}
\input{Existing}

\section{Future Directions}
\input{Future}

\section{Conclusion}
\input{Conclusion}

\bibliographystyle{named}
\bibliography{ijcai23}

\end{document}

%% file: Introduction.tex
The rapid development of AI chipsets has seen emerging mobile devices become ubiquitous platforms, such as smartphones, robotics and autonomous driving systems. A broad range of capabilities are available for these devices and vast amounts of data can be collected from end users. Leveraging these data to enhance AI model learning and improve the user experience poses tremendous potential. As a result, traditional compute-intensive AI-based tasks have been migrating to mobile devices~\cite{AItask_1,AItask_2,AItask_3,AItask_4}, which facilitates the deployment of high-performance AI applications in open-world environments.

However, with the advent of increasingly strict privacy regulations ~\cite{policy_DMF,policy_CCPA,policy_GDPR}, it has become more challenging to collect and utilize user data on the centralized servers to build AI models. Executing AI tasks on local devices is an effective approach to solve this issue, as it contributes to prevent local data leakage. Meanwhile, training AI tasks locally also benefits from multiple options, including reduced latency, lower power consumption, and decreased bandwidth requirements. Consequently, exploring how mobile devices can perform AI tasks locally has emerged as a prominent research topic within the field.

\textbf{Despite the promising benefits, how to perform on-device training still remains challenges.} First, edge devices are memory-constrained.
Training deep learning models on-device poses challenges due to memory scarcity on-device processors. They are significantly inefficient and have less available memory compared to server-based processors, making it the primary challenge. Furthermore, the rapid progress in deep learning has sparked the emergence of ever-growing neural network models, often referred to as the ``model scaling wars", especially in the era of Artificial Intelligence Generation and Consumption (AIGC). Meanwhile, training is a long-term task that is highly sensitive to both computational resources and memory. Unlike inference, the training process necessitates the storage of all intermediate activations, in addition to the need to retain model states and weights throughout the iterative iterations. For instance, consider MobileNetV2~\cite{mobilenetv2}, a widely employed multi-task model for on-device applications. It demands approximately 8GB of memory (with a batch size of 32) during the training process, whereas only around 700MB of memory suffices for inference. This striking disparity presents a substantial hurdle when it comes to training large-scale models on mobile devices, which typically possess limited RAM ranging from 4-12GB~\cite{RAM,Sage}.

The combination of high memory requirements and limited resources poses a significant challenge that prevents many mobile devices from participating in the learning process. This constraint in available memory, commonly known as the ``memory wall", can result in Out-Of-Memory (OOM) errors and impede the deployment of resource-intensive training tasks on mobile devices. As a result, the data stored on these devices cannot be effectively utilized for training purposes.
To overcome this limitation and enable effective and scalable deep learning model training on resource-constrained devices, it is crucial to address the memory constraints and develop memory-efficient strategies. By optimizing memory usage and adopting techniques that minimize memory footprint, it becomes possible to leverage the potential of these devices for training complex models. This, in turn, allows for efficient utilization of data on mobile devices and expands the possibilities of on-device learning.
By tackling the memory constraints and developing memory-efficient strategies, the barriers to mobile device participation in the learning process can be significantly reduced. This opens up opportunities for broader and more inclusive utilization of deep learning models, empowering resource-constrained devices to contribute effectively to the training process.

\textbf{Outline.}
The primary objective of this survey is to provide a comprehensive summary of the methods that have been developed to address the memory wall in on-device learning. The survey encompasses both hardware performance improvements and algorithmic optimizations, considering the diverse approaches that have been implemented.
These methods showcase innovative strategies aimed at leveraging available resources more effectively, thereby mitigating the memory limitations in on-device learning. By exploring and incorporating advancements in both hardware and algorithms, researchers and practitioners have made significant progress in improving the efficiency of on-device learning.

However, it is essential to acknowledge that despite these advancements, there is still a substantial gap to bridge in breaking down the memory walls and further enhancing the efficiency of on-device learning. The complexities and challenges associated with resource-constrained environments necessitate continuous research and development efforts. Future work should continue to explore novel techniques, refine existing approaches, and push the boundaries to overcome the memory limitations and achieve more efficient on-device learning systems.
By recognizing the current progress, as well as the ongoing challenges and opportunities, this survey aims to contribute to the collective understanding of the field and inspire further research in the pursuit of efficient on-device learning.

The remaining sections of the paper are structured as follows. Section 2 provides a detailed explanation of the factors contributing to the memory wall phenomenon encountered during on-device training. In Section 3, existing solutions to address this challenge are analyzed and discussed. Following this, Section 4 presents an outlook on the potential synergy between on-device training frameworks and other training approaches. Finally, the conclusions of the study are presented in the concluding section.

%% file: MemoryWall.tex
The recent advancements in mobile system-on-chip (SoC) technology and software have sparked innovative research on pushing the boundaries of deep learning model utilization on resource-limited platforms. With the ability to perform inference on relatively heavy models directly on devices, the training of simplified versions of these models can now be done locally. However, despite the efforts made by the machine learning research community to develop more efficient neural networks, the field of mobile computing has yet to address the challenge of training and customizing state-of-the-art heavy and complex models on mobile platforms. This gap in current research highlights the need for further exploration to enable effective training and customization of these advanced models on mobile devices.

\subsection{Performance gap between device-based and server-based processors}

Although modern devices are equipped with dedicated accelerators such as NPUs and GPUs to enhance neural network processing which application processor (AP) utilizes similar compute elements as a server-side one, but operates at lower power consumption and occupies a smaller physical space.
there remains a significant performance gap compared to server-based implementations that leverage powerful hosts and top-of-the-line processors. This gap is primarily due to the architectural differences between smartphones and server-based one.

Modern smartphones are composed of a diverse collection of specialized cores like CPU, GPU, memory, and other heterogeneous components following a SoCs architecture, which work together to perform various tasks. While these accelerators provide valuable processing capabilities, training deep neural networks on smartphone SoCs without careful consideration of their resource constraints can lead to suboptimal training performance and significantly impact the user experience.

\begin{table}[htbp]
  \centering
  \captionsetup{justification=centering, labelsep=period}
  \caption{Queries Per Second (QPS) for Different Processors.}
  \begin{tabular}{ccc}
    \toprule
    \multicolumn{2}{c}{\textbf{Processor}} & \textbf{QPS} \\
    \cmidrule(r){1-2} \cmidrule(l){3-3}
    \textbf{Type} & \textbf{Model} & \\
    \midrule
    \multirow{3}{*}{Server-based} & NVIDIA A100-SXM-80GB & 25403.7 \\
    & NVIDIA Quadro RTX 6000 & 10653 \\
    & NVIDIA Quadro RTX 8000 & 10804 \\
    \midrule
    \multirow{6}{*}{Device-based} & Qualcomm Snapdragon 8 Gen 2 & 197.99 \\
    & Snapdragon 7 Gen 1 & 102.69 \\
    & Exynos 2300 & 109.21 \\
    & Snapdragon 888+ & 24.17 \\
    & Snapdragon 865+ & 6.29 \\
    & Dimensity 820 & 3.85 \\
    \bottomrule
  \end{tabular}
  \label{tab: device}
\end{table}

In contrast, deep learning GPUs designed for servers have made remarkable advancements, enabling efficient execution of complex operations in large batches and facilitating the utilization of sophisticated deep learning models. However, despite improvements in mobile GPUs, they still lack the processing power and peripheral resources available on server-side counterparts. As a result, system designers are faced with the challenge of compromising on performance, as fully exploiting complex deep learning models, including training operations, is currently not feasible on mobile platforms.
Table~\ref{tab: device} displays the QPS (Queries Per Second) of different processors when performing natural language processing inference tasks using the MobileBERT~\cite{mobilebert}, a language model with 25 million parameters, on the SQuAD dataset. 
The Qualcomm Snapdragon 8 Gen 2, released in 2022, achieves a processing rate of 197.99 Queries Per Second (QPS) ~\cite{A100}. In comparison, a server-based NVIDIA A100-SXM-80GB GPU, also released in 2022, achieves a significantly higher rate of 25,403.70 QPS~\cite{A100}. This stark contrast in performance highlights the scarcity of processing power on mobile GPUs.

It is worth noting that the performance gap may vary depending on the GPU models used. However, it can be reasonably concluded that mobile GPUs currently offer limited processing power compared to their server-based counterparts.

\begin{principle-1}
On-device processor capability hinders model training.
\end{principle-1}

\subsection{Models grow increasingly complicated}
Accuracy improvements in machine learning-based solutions are largely driven by scale, as previous researches have indicated ~\cite{model_1,olla}. One significant aspect contributing to this improvement is the expansion of both the depth and width of DNNs. These dimensions of DNNs expand exponentially, reflecting the increasing complexity and capacity of these models. 
As a result, the memory requirements for storing network weights and intermediate results such as activations and gradients during the training process also grow. This inflation in size poses a challenge, as more memory resources are necessary to accommodate the expanding parameters and data. 

Furthermore, with the increasing complexity of application scenarios, the design of ``big models" has become necessary to achieve optimal performance in such scenarios. Researchers are now training neural networks on larger inputs, such as high-resolution images or longer sequences of data, which further exacerbates the memory demands.
As an example, the 50-layer ResNet network~\cite{AItask_5} consists of 26 million weight parameters and performs 16 million activations during the forward pass. Storing each weight and activation as a 32-bit floating-point value would require a total storage capacity of 168 MB.
While the situation worsens with the introduction of GPT-3.5 (Generative Pre-trained Transformer) , which incorporates multi-layer transformers with its 170 billion parameters, further exacerbates the storage requirements and computational complexity associated with such large models. Therefore, the combination of larger network sizes and inputs intensifies the memory requirements for training deep neural networks.

\begin{principle-2}
``Model scaling wars" causing memory footprint limits on device training.
\end{principle-2}

\subsection{Memory footprint gap between on-devive training and inference}

While the mobile machine learning system stack continues to advance, there are ongoing advancements on algorithm optimizations and neural network architectures taking place. Furthermore, the emergence of high-performance computing AI chips, such as programmable accelerators, are further driving the optimization of on-device inference efficiency.
However, the same optimizations cannot be carried out for the ML model training process.
\begin{figure}[ht]
    \centering
    \includegraphics[width=1.0\linewidth]{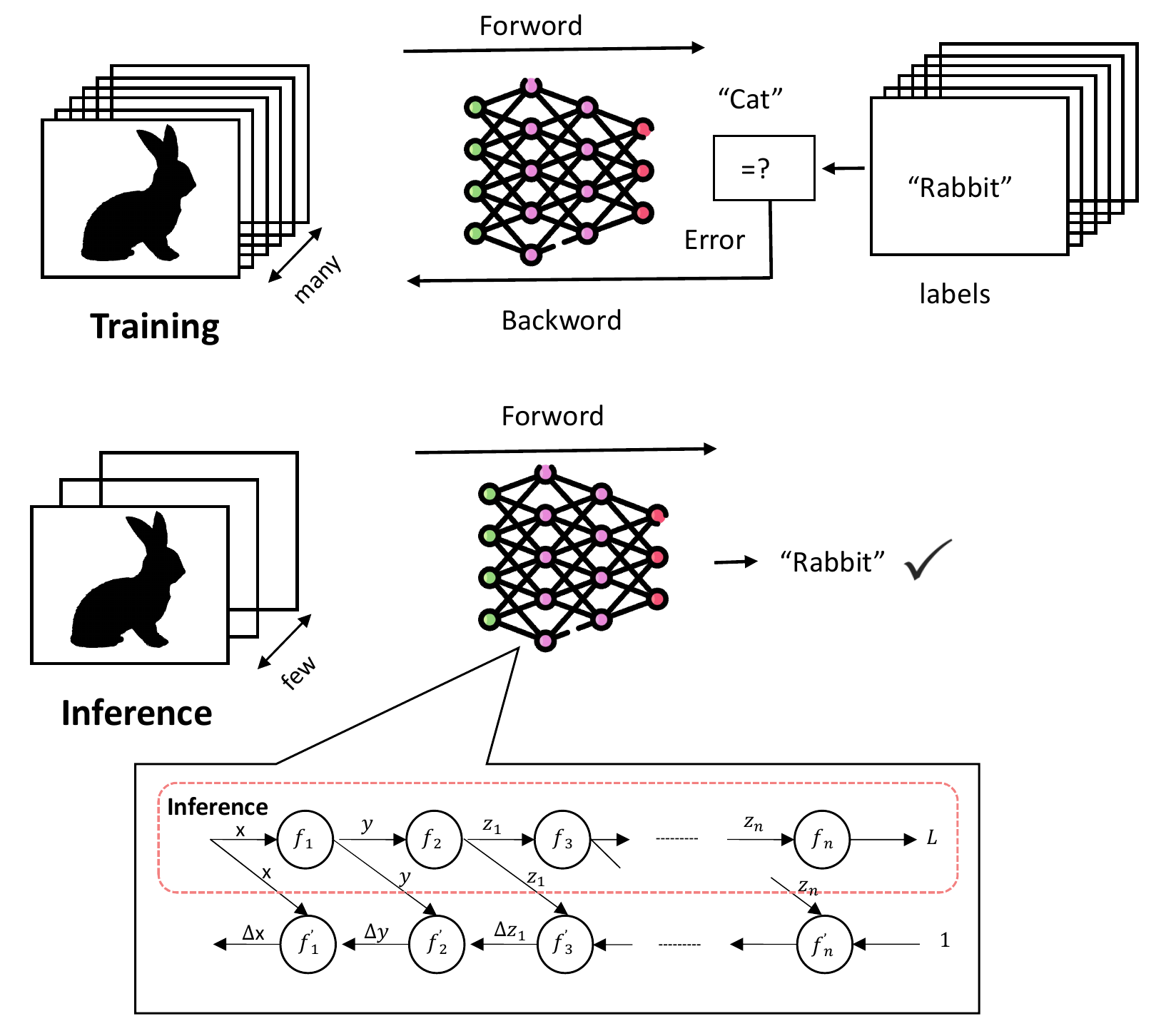}
    \caption{Model training v.s. inference.}
    \label{fig:train}
\end{figure}

Figure~\ref{fig:train} illustrates the computational graph representation of the ML model inference and training process~\cite{Sage}.
Within a neural network layer, which consists of an input vector $\vec{x}$ and an output vector $\vec{y}$, the loss function denoted as $L=g(\vec{y})$ is utilized. The gradient $\nabla \vec{x}=\left(\frac{\partial L}{\partial x_1}, \ldots, \frac{\partial L}{\partial x_n}\right)$ is computed using the chain rule:

\begin{equation}
\frac{\partial L}{\partial x_n}=\frac{\partial L}{\partial y_n} \times \frac{\partial y_n}{\partial x_n}
\end{equation}

This equation highlights the dependence of the gradient $\nabla \vec{x}$ on both the output gradient $\nabla \vec{y}$ and the original input $\vec{x}$. The computational graph during model inference is straightforward, with intermediate states being discardable after use. However, in the context of model training, the graph topology becomes more complex. Each gradient node $\nabla \vec{x}$ relies on both its output gradient $\nabla \vec{y}$ and the original input value $\vec{x}$ due to automatic differentiation principles. Consequently, the forward process generates a significant amount of data that must be stored in the memory buffer for back-propagation gradient generation. Thus, the training process necessitates a considerably larger memory footprint compared to inference operations, often requiring multiple GBs of memory on platforms with limited memory capacity.

\begin{figure}
    \centering
    \includegraphics[width=1.0\linewidth]{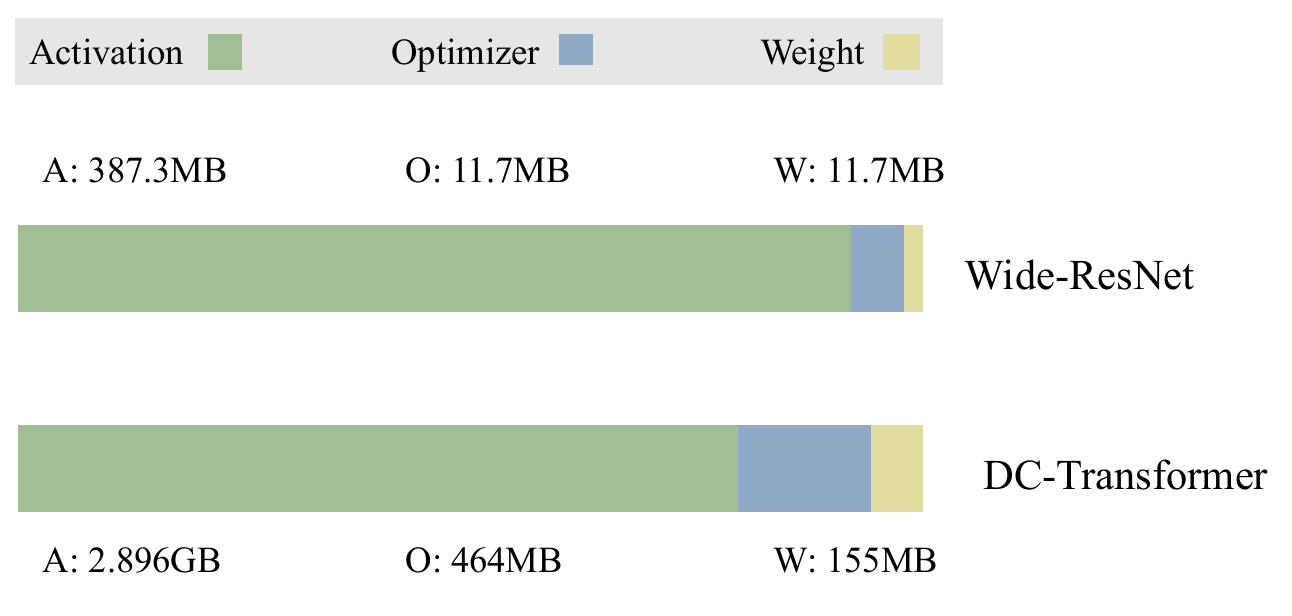}
    \caption{The memory consumption breakdown during DNN
training.}
    \label{fig:memory}
\end{figure}

Furthermore, the training overhead is further magnified due to the extended duration of training tasks. Memory usage during training can be 5 to 100 times greater than during inference operations, depending on factors such as input size and batch size. For on-device training (refer to Figure~\ref{fig:memory} ~\cite{sohoni2019low}), the total memory consumption can be attributed to three types of memory: weight, optimizer, and activation.

\begin{principle-3}
Training is considerably more overhead than inference, and training memory is dominated by activation rather than parameters.
\end{principle-3}

Therefore, the memory requirement plays a vital role in the scalability of deep learning training due to its resource-intensive nature, which necessitates considerably more hardware resources compared to inference operations. The larger memory footprint and training overhead present challenges when deploying models for local training.

%% file: Existing.tex
To facilitate the memory-efficient deployment of model training on mobile devices, three distinct approaches have been proposed: hardware-based, model-based, and process-based. The hardware-based approach concentrates on optimizing memory utilization at the physical level, starting from scratch. The model-based approach aims to enhance the efficiency of model parameters to reduce the memory requirements for training. However, this approach may entail a trade-off in model performance. On the other hand, the process-based approach involves adjusting the training paradigm to decrease the per-round training memory. Nevertheless, this adjustment may lead to longer training times and inefficiencies. In the subsequent sections, we will analyze these approaches from three different perspectives.

\subsection{Hardware enable memory-friendly}
\textbf{Device Training Libraries.}
The significant memory consumption of mobile operating systems and their background applications poses a significant challenge for training complex models on mobile platforms. This limitation further complicates the task of training such models. However, the increased complexity of these models leads to improved accuracy and expands their potential applications. Consequently, it becomes essential to develop a framework that effectively manages memory usage during DNN training on mobile platforms. 
Encouragingly, there has been progress in the form of an implementation framework specifically tailored for mobile devices. These efforts aim to efficiently address the memory requirements of DNN training, ensuring smooth execution on mobile platforms.

 Libraries like Apple's CoreML~\cite{coreml} for iOS devices and Android's Neural Networks API (NNAPI)~\cite{nnapi} for Android devices enable offloading model inference to the mobile GPU or NPU, resulting in accelerated inference performance. These libraries offer convenient interfaces and optimizations for efficient inference on mobile platforms. While there are various libraries available to support model inference on mobile devices, the options for on-device training frameworks are relatively scarce. Our research has identified a few libraries that can potentially be used for on-device training, namely Deeplearning4J~\cite{deeplearningforjava}, MNN~\cite{MNN}, TFLite~\cite{tflite} and PyTorch provide Java\cite{Pytorch}. These libraries provide Java or Python interfaces that can be integrated into mobile devices. However, it should be noted that these frameworks are not optimized for space efficiency and can occupy a significant amount of memory, sometimes reaching up to 400 MB~\cite{Swan}. Moreover, existing mobile engines lack the capability to offload training tasks to GPUs, which further restricts their utility for on-device training. Furthermore, these mobile engines typically necessitate significant engineering efforts to implement and experiment with new designs, rendering the development process more intricate and time-consuming. Hence, there is a demand for more efficient and user-friendly frameworks that are specifically tailored to facilitate on-device training on mobile platforms.

\textbf{Hardware Optimization.}
Indeed, optimizing hardware for memory is a direct approach to address the memory constraints encountered in on-device training. By enhancing the design and architecture of hardware components, it becomes feasible to improve the performance, efficiency, and capacity of the memory subsystem, thereby mitigating the limitations posed by memory constraints. 

\textit{Leveraging Idle Resources for Memory Optimization} focuses on utilizing idle resources, such as GPU idle cycles or monitoring idle times in cloud gaming, to optimize memory usage. 
PilotFish~\cite{PilotFish} introduces a method for utilizing the idle cycles in cloud gaming for deep learning training. By scheduling training computation kernels during these idle cycles, the GPU's potential is effectively harnessed, leading to optimized workload allocation, parallelism, and reduced memory usage.
Swan~\cite{Swan} presents a neural engine for efficient DNN training on smartphone SoCs. It employs a three-part approach consisting of monitoring, exploring execution choices, and on-device training. This approach optimizes performance, energy efficiency, and memory usage, resulting in faster training times and improved memory efficiency.

\textit{Hardware Resource Utilization for Memory Pressure Reduction} emphasizes hardware resource utilization, including CPU-DSP co-scheduling, memory pool optimization, and integrated search space optimization, to alleviate memory pressure. 
Mandheling~\cite{Mandheling} proposes a mixed-precision on-device DNN training method with CPU-DSP co-scheduling. By overlapping CPU and DSP execution and minimizing context switchings, the overhead of DSP-unfriendly operators is reduced. This optimization technique enhances resource utilization and helps mitigate memory pressure.
Melon~\cite{Melon} aims to overcome the memory limitation of mobile devices when training with large batch sizes. It optimizes memory pool utilization based on the MNN framework, enabling efficient training with larger batch sizes. This approach reduces memory pressure and improves overall resource efficiency.
POET~\cite{Poet} addresses memory consumption in backpropagation by optimizing the integrated search spaces of rematerialization and paging algorithms. The goal is to achieve efficient memory management and minimize the overall memory footprint of the backpropagation algorithm. This optimization technique effectively reduces memory pressure during training.

By focusing on hardware optimization of memory, it is possible to enhance the memory capabilities of on-device training systems, enabling more efficient and effective execution of training tasks within the constraints of mobile and embedded devices.

\begin{limitationbox}
    However, it is noted that most of these methods are not directly focused on memory optimization or have high requirements for aspects other than hardware, while hardware optimization produces little gain. For example, Melon offers only small gains in memory pool optimization, limiting its broader applicability.
\end{limitationbox}

\subsection{Model performance and memory trade-off}
Besides the hardware approaches, lowering the model performance requirements via decreasing the model's computational complexity is another way to reduce the training memory that involve restructuring the model through techniques such as quantization~\cite{Octo,int8}, sparse~\cite{sparse,sparse_2}, pruning~\cite{pruning} to minimize memory consumption. 

\begin{table}[htbp]
  \centering
  \caption{Optimization Techniques for Training Memory Efficiency.}
  \resizebox{\linewidth}{!}{%
    \begin{tabular}{@{}cccccc@{}}
    \toprule
    \multicolumn{2}{c}{Memory} & Weight & Optimizer & Activations \\
    \midrule
    \multirow{3}{*}{Model Performance} & Quantization & $\downarrow$ & $\downarrow$ & $\downarrow$ \\
    \cmidrule{2-5} 
          & Sparse & $\downarrow$ & $\downarrow$ & \\
    \cmidrule{2-5} 
          & Pruning & $\downarrow$ & $\downarrow$ & \\
    \midrule
    \multirow{3}{*}{Training Efficiency} & Micro-batch & & & $\downarrow$ \\
    \cmidrule{2-5} 
          & Checkpoint & & & $\downarrow$ \\ 
    \bottomrule
    \end{tabular}%
  }
  \label{tab:addlabel}%
\end{table}%

\textit{Quantization} aims to reduce memory usage by representing model parameters and activations with lower precision.
Octo~\cite{Octo} is a lightweight, cross-platform system for tiny on-device learning that addresses the challenge of reducing memory overhead. It introduces INT8 training with loss-aware compensation and backward quantization, enabling efficient training on resource-constrained devices. By utilizing a loss-aware compensation technique and backward quantization, Octo minimizes the memory footprint while preserving model accuracy.
Additionally, the use of lower-precision floating points or integer units, as discussed in~\cite{mixed}, further reduces the computational and storage costs associated with model operations. This approach leverages the benefits of lower-precision representations to optimize memory utilization and computational efficiency.

\textit{Sparsity} leverages the inherent sparsity in deep learning computations to reduce memory overhead.
 ~\cite{sparse} introduces sparse GPU kernels specifically designed for deep learning tasks. By leveraging sparsity within GPU kernels, this approach effectively reduces memory overhead and provides efficient computation for deep learning workloads.
~\cite{sparse_2} presents a dynamic sparse graph approach for efficient deep learning. By dynamically constructing sparse computational graphs, it reduces memory usage and enables efficient deep learning computations.

\textit{Pruning} selectively removes unnecessary model parameters to reduce memory usage while maintaining model performance. ~\cite{pruning} proposes filter pruning based on the geometric median for accelerating deep convolutional neural networks. By pruning filters, it reduces the number of model parameters, thereby reducing memory overhead and improving computational efficiency.

\begin{limitationbox}
However, these techniques may impair the model structure and lead to a degradation of the model's performance, resulting in lower accuracy.
\end{limitationbox}

\subsection{Training efficiency and memory trade-off}
As mentioned before, activations play a crucial role in the memory footprint during the training process of deep learning models. Recomputing discarded activations~\cite{activation_1,activation_2} during the backward pass, employing microbatching\cite{sohoni2019low,gpipe} and gradient checkpointing~\cite{Sage} are effective techniques to reduce memory usage without compromising the training process.

\textit{Gradient Checkpointing} selectively stores and recomputes intermediate values during the backward pass. By trading computation time for memory usage, it enables more efficient training on memory-constrained devices. Sage~\cite{Sage} optimizes the memory usage in DNN gradient evaluation through a flexible computation graph configuration and a combination of operator and graph-level optimizations. During runtime, it dynamically adjusts its memory consumption using a hybrid approach of gradient checkpointing and micro-batching techniques, based on the available system memory budget. Firstly, Sage constructs a computation graph using automatic differentiation (AD), with gradients represented as separate nodes. This allows for the implementation of graph- and operator-level optimizations, reducing the static memory footprint. Finally, during the gradient descent computation, Sage adaptively combines gradient checkpointing and gradient accumulation, taking into account the system's memory capacity and computational capabilities. This adaptive strategy enables efficient memory utilization during the graph evaluation process.Melon~\cite{Melon} addresses the challenge of breaking the memory wall for resource-efficient on-device machine learning. One of the key methods it proposes is the use of recomputing to reduce training memory overhead. By optimizing memory pool utilization based on the MNN framework, Melon enables training tasks with large batch sizes that exceed the physical memory capacity of mobile devices. It achieves this by recomputing intermediate values during the training process, allowing for efficient memory utilization and reducing the overall memory footprint. This innovative approach significantly alleviates the memory constraints in on-device machine learning, enabling more resource-efficient training on mobile devices.

\textit{Microbatching} involves dividing the training data into smaller subsets called microbatches. Processing these microbatches sequentially reduces memory usage compared to processing the entire dataset in one batch. GPipe~\cite{gpipe} introduces microbatching as a memory-saving technique for efficient training of large neural networks using pipeline parallelism. By breaking down the training process into microbatches, GPipe reduces memory requirements during model forward and backward passes. Instead of processing the entire batch at once, microbatching allows for incremental storage and release of intermediate activations, significantly reducing the overall memory footprint. This memory-saving approach enables training of large-scale models with limited computational resources, making it valuable for reducing training memory overhead.

Indeed, some of the approaches mentioned earlier may reduce memory usage but come at the expense of a large computation overhead. This trade-off makes them less preferred for edge devices, which typically have limited computational resources and prioritize efficiency. 

\begin{limitationbox}
    However, it is equally important to consider the computational requirements and efficiency of these approaches. A solution that reduces memory usage but significantly increases computation overhead may not be practical or suitable for edge devices.
\end{limitationbox}

%% file: Future.tex
Training models on mobile devices presents challenges, but it offers several advantages. By enabling applications to train deep neural networks locally, it promotes a better understanding of and safeguards privacy-sensitive data. Additionally, it reduces dependence on infrastructure, allowing users to provide services in diverse external environments. This approach also incorporates the advancements made in federated learning research. Furthermore, on-device training facilitates fine-tuning of the initially provided generic deep learning model, enabling efficient updates to model parameters. Leveraging locally collected data directly and seamlessly for parameter updates has the potential to accelerate mobile computing.

Although existing ML algorithms have made considerable progress to train the model on the edge, but there are still challenges and limitations that need to be addressed to further optimize and improve ML on the edge.

\begin{figure}[!ht]
    \centering
    \includegraphics[width=1 \linewidth]{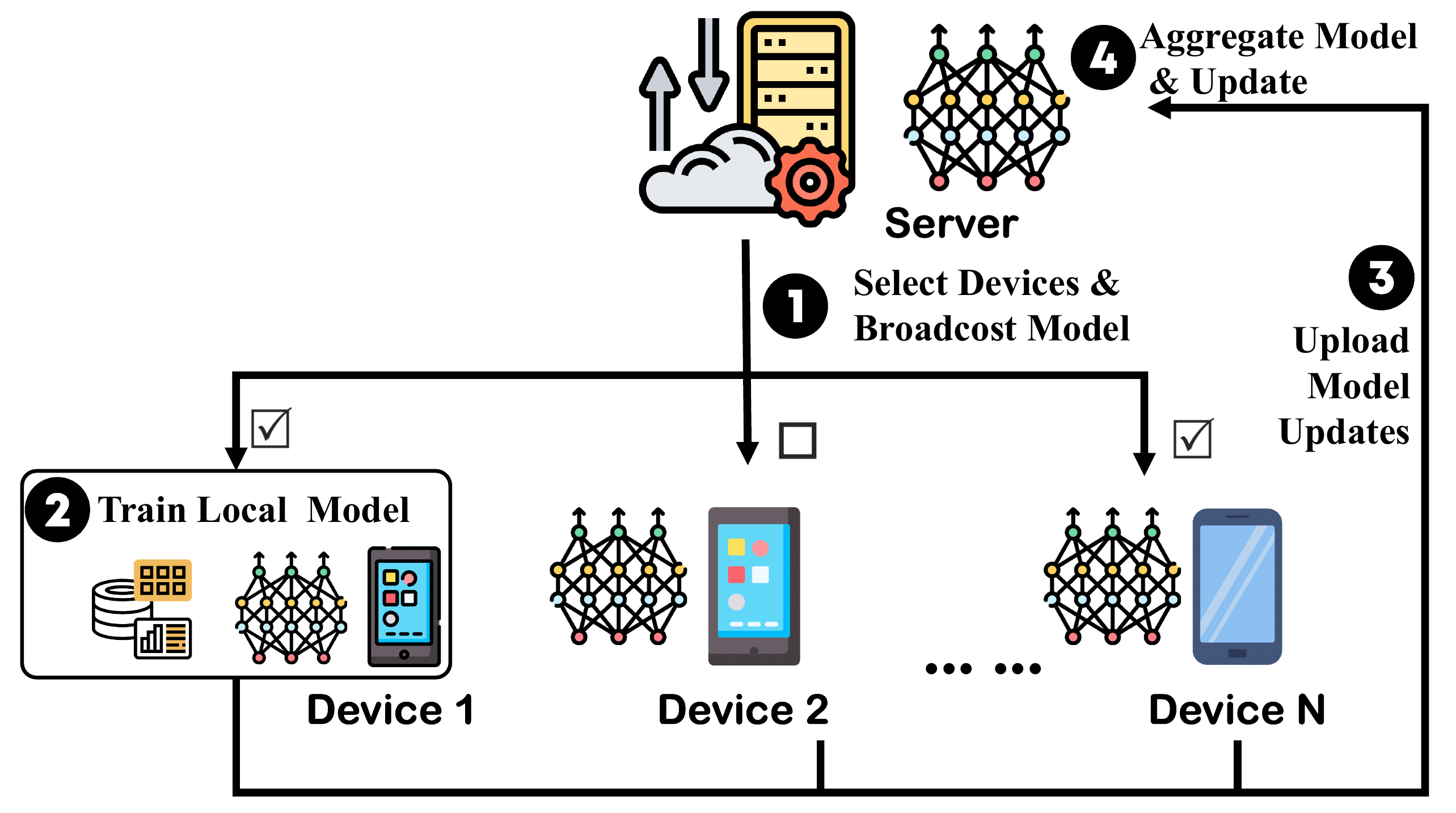}
    \vspace{-1em}
    \caption{Workflow of Federated Learning.}
    \label{fig:FL}
    \vspace{-1em}
\end{figure}

Recently, federated learning~\cite{2016federated,fedavg} (FL) as a popular distributed learning paradigm enable scalable intelligence with private and large data.
FL collaboratively trains a shared model across multiple devices, where each device is trained in parallel using its local data for privacy awareness. A holistic framework of representative FL is shown in Figure~\ref{fig:FL}. \ding{182} The server randomly selects mobile devices from the device pool for each training round; \ding{183} then the server delivers the initialized collaborative models to the selected devices;\ding{184} With the received model, local devices train in parallel using its own data for a specified round; \ding{185} While local devices upload the model updates to the central server; \ding{186} There, the server aggregates the obtained models to update the collaborative model. The training process iterates over and over until the model converges or achieves the ideal accuracy.

Some efforts also exploit the inter-data benefits in the federated learning framework to perform memory optimization during the on-device training process. HeteroFL\cite{heterofl} allows for model architecture heterogeneity among participating clients by adjusting model widths based on the processing power of each client. This approach tailors the model size to the available resources of each device, thereby optimizing memory usage during training. Split-Mix\cite{efficient} splits an extensive neural network into several base sub-networks to reduce the training process footprint. Fedepth~\cite{zhang2023memory} adaptively decomposes the full model into blocks based on the memory budgets of individual clients. It trains these blocks sequentially to obtain a full inference model. This approach dynamically adjusts the model size according to each client's memory constraints, enabling efficient memory utilization in FL scenarios. Also, the system and data heterogeneity~\cite{tian2022harmony} across multiple devices due to distributed training is an urgent problem to be solved. These approaches aim to break the memory limitations on edge devices while ensuring training efficiency and model performance. However, insufficient model training parameters and the inability to train a complete network usually have a damaging effect on the holistic performance. They leverage techniques such as model heterogeneity, model decomposition, and adaptive memory management to optimize memory usage during on-device training within the federated learning framework. Thus, a framework that break edge memory wall and ensure the training efficiency and model performance is urgently required.

%% file: Conclusion.tex
In this survey, we systematically explore the methods devised to overcome memory limitations in on-device learning. By conducting a comprehensive review of the existing literature, we provide an overview of state-of-the-art solutions to the memory wall problem. We analyze a diverse range of approaches, aiming to offer a comprehensive understanding of how researchers have addressed memory constraints in on-device learning. Additionally, we identify and discuss open problems that require further attention in this field. These challenges present opportunities for advancements in memory optimization for on-device learning. We propose potential research directions to overcome these challenges and drive the field forward. Our survey contributes to the knowledge base, consolidates existing research efforts, and guides future endeavors in tackling memory limitations in on-device learning.

%% file: main.bbl
\begin{thebibliography}{}

\bibitem[\protect\citeauthoryear{Abadi \bgroup \em et al.\egroup
  }{2019}]{tflite}
Martín Abadi, Paul Barham, Jianmin Chen, Zhifeng Chen, Andy Davis, Jeffrey
  Dean, Matthieu Devin, Sanjay Ghemawat, Geoffrey Irving, Michael Isard, et~al.
\newblock Tensorflow lite: Ml for mobile and edge devices, 2019.

\bibitem[\protect\citeauthoryear{{ASEAN}}{2021}]{policy_DMF}
{ASEAN}.
\newblock Asean data management framework.
\newblock
  \url{https://asean.org/wp-content/uploads/2-ASEAN-Data-Management-Framework_Final.pdf},
  2021.

\bibitem[\protect\citeauthoryear{Bubeck and Sellke}{2023}]{model_1}
S{\'e}bastien Bubeck and Mark Sellke.
\newblock A universal law of robustness via isoperimetry.
\newblock {\em Journal of the ACM}, 70(2):1--18, 2023.

\bibitem[\protect\citeauthoryear{Chen \bgroup \em et al.\egroup
  }{2016}]{activation_1}
Tianqi Chen, Bing Xu, Chiyuan Zhang, and Carlos Guestrin.
\newblock Training deep nets with sublinear memory cost.
\newblock {\em arXiv preprint arXiv:1604.06174}, 2016.

\bibitem[\protect\citeauthoryear{de~la Torre}{2018}]{policy_CCPA}
Lydia de~la Torre.
\newblock A guide to the california consumer privacy act of 2018.
\newblock {\em Available at SSRN 3275571}, 2018.

\bibitem[\protect\citeauthoryear{Diao \bgroup \em et al.\egroup
  }{2020}]{heterofl}
Enmao Diao, Jie Ding, and Vahid Tarokh.
\newblock Heterofl: Computation and communication efficient federated learning
  for heterogeneous clients.
\newblock {\em arXiv preprint arXiv:2010.01264}, 2020.

\bibitem[\protect\citeauthoryear{Eklund \bgroup \em et al.\egroup
  }{2017}]{deeplearningforjava}
Adam Eklund, Stephen Merity, Daniel Blake, and Suvrat Ravi.
\newblock Deeplearning4j: Open-source deep learning for the jvm.
\newblock {\em arXiv preprint arXiv:1711.05413}, 2017.

\bibitem[\protect\citeauthoryear{Fang \bgroup \em et al.\egroup
  }{2018}]{AItask_2}
Biyi Fang, Xiao Zeng, and Mi~Zhang.
\newblock Nestdnn: Resource-aware multi-tenant on-device deep learning for
  continuous mobile vision.
\newblock In {\em Proceedings of the 24th Annual International Conference on
  Mobile Computing and Networking}, pages 115--127, 2018.

\bibitem[\protect\citeauthoryear{Fang \bgroup \em et al.\egroup
  }{2020}]{AItask_1}
Biyi Fang, Xiao Zeng, Faen Zhang, Hui Xu, and Mi~Zhang.
\newblock Flexdnn: Input-adaptive on-device deep learning for efficient mobile
  vision.
\newblock In {\em 2020 IEEE/ACM Symposium on Edge Computing (SEC)}, pages
  84--95. IEEE, 2020.

\bibitem[\protect\citeauthoryear{Gale \bgroup \em et al.\egroup
  }{2020}]{sparse}
Trevor Gale, Matei Zaharia, Cliff Young, and Erich Elsen.
\newblock Sparse gpu kernels for deep learning.
\newblock In {\em SC20: International Conference for High Performance
  Computing, Networking, Storage and Analysis}, pages 1--14. IEEE, 2020.

\bibitem[\protect\citeauthoryear{Gim and Ko}{2022}]{Sage}
In~Gim and JeongGil Ko.
\newblock Memory-efficient dnn training on mobile devices.
\newblock In {\em Proceedings of the 20th Annual International Conference on
  Mobile Systems, Applications and Services}, pages 464--476, 2022.

\bibitem[\protect\citeauthoryear{Greff \bgroup \em et al.\egroup
  }{2016}]{activation_2}
Klaus Greff, Rupesh~K Srivastava, and J{\"u}rgen Schmidhuber.
\newblock Highway and residual networks learn unrolled iterative estimation.
\newblock {\em arXiv preprint arXiv:1612.07771}, 2016.

\bibitem[\protect\citeauthoryear{He \bgroup \em et al.\egroup
  }{2016}]{AItask_5}
Kaiming He, Xiangyu Zhang, Shaoqing Ren, and Jian Sun.
\newblock Deep residual learning for image recognition.
\newblock In {\em Proceedings of the IEEE conference on computer vision and
  pattern recognition}, pages 770--778, 2016.

\bibitem[\protect\citeauthoryear{He \bgroup \em et al.\egroup }{2020}]{pruning}
Yihui He, Jian Lin, Zhuang Liu, Hanrui Wang, Liang Li, and Song Han.
\newblock Filter pruning via geometric median for deep convolutional neural
  networks acceleration.
\newblock In {\em Proceedings of the IEEE/CVF Conference on Computer Vision and
  Pattern Recognition}, pages 4340--4349, 2020.

\bibitem[\protect\citeauthoryear{Hong \bgroup \em et al.\egroup
  }{2022}]{efficient}
Junyuan Hong, Haotao Wang, Zhangyang Wang, and Jiayu Zhou.
\newblock Efficient split-mix federated learning for on-demand and in-situ
  customization.
\newblock {\em arXiv preprint arXiv:2203.09747}, 2022.

\bibitem[\protect\citeauthoryear{Huang \bgroup \em et al.\egroup
  }{2019}]{gpipe}
Yanping Huang, Youlong Cheng, Ankur Bapna, Orhan Firat, Dehao Chen, Mia Chen,
  HyoukJoong Lee, Jiquan Ngiam, Quoc~V Le, Yonghui Wu, et~al.
\newblock Gpipe: Efficient training of giant neural networks using pipeline
  parallelism.
\newblock {\em Advances in neural information processing systems}, 32, 2019.

\bibitem[\protect\citeauthoryear{Inc.}{2017}]{coreml}
Apple Inc.
\newblock Core ml: Machine learning for ios.
\newblock In {\em Apple Developer Documentation}. Apple Inc., 2017.

\bibitem[\protect\citeauthoryear{Jiang \bgroup \em et al.\egroup
  }{2020}]{AItask_3}
Shuang Jiang, Zhiyao Ma, Xiao Zeng, Chenren Xu, Mi~Zhang, Chen Zhang, and
  Yunxin Liu.
\newblock Scylla: Qoe-aware continuous mobile vision with fpga-based dynamic
  deep neural network reconfiguration.
\newblock In {\em IEEE INFOCOM 2020-IEEE Conference on Computer
  Communications}, pages 1369--1378. IEEE, 2020.

\bibitem[\protect\citeauthoryear{Kone{\v{c}}n{\`y} \bgroup \em et al.\egroup
  }{2016}]{2016federated}
Jakub Kone{\v{c}}n{\`y}, H~Brendan McMahan, Felix~X Yu, Peter Richt{\'a}rik,
  Ananda~Theertha Suresh, and Dave Bacon.
\newblock Federated learning: Strategies for improving communication
  efficiency.
\newblock {\em arXiv preprint arXiv:1610.05492}, 2016.

\bibitem[\protect\citeauthoryear{Liu \bgroup \em et al.\egroup
  }{2018}]{sparse_2}
Liu Liu, Lei Deng, Xing Hu, Maohua Zhu, Guoqi Li, Yufei Ding, and Yuan Xie.
\newblock Dynamic sparse graph for efficient deep learning.
\newblock {\em arXiv preprint arXiv:1810.00859}, 2018.

\bibitem[\protect\citeauthoryear{Lv \bgroup \em et al.\egroup }{2022}]{MNN}
Chengfei Lv, Chaoyue Niu, Renjie Gu, Xiaotang Jiang, Zhaode Wang, Bin Liu, Ziqi
  Wu, Qiulin Yao, Congyu Huang, Panos Huang, et~al.
\newblock Walle: An $\{$End-to-End$\}$,$\{$General-Purpose$\}$, and
  $\{$Large-Scale$\}$ production system for $\{$Device-Cloud$\}$ collaborative
  machine learning.
\newblock In {\em 16th USENIX Symposium on Operating Systems Design and
  Implementation (OSDI 22)}, pages 249--265, 2022.

\bibitem[\protect\citeauthoryear{McMahan \bgroup \em et al.\egroup
  }{2017}]{fedavg}
Brendan McMahan, Eider Moore, Daniel Ramage, Seth Hampson, and Blaise~Aguera
  y~Arcas.
\newblock Communication-efficient learning of deep networks from decentralized
  data.
\newblock In {\em Artificial intelligence and statistics}, pages 1273--1282.
  PMLR, 2017.

\bibitem[\protect\citeauthoryear{Micikevicius \bgroup \em et al.\egroup
  }{2017}]{mixed}
Paulius Micikevicius, Sharan Narang, Jonah Alben, Gregory Diamos, Erich Elsen,
  David Garcia, Boris Ginsburg, Michael Houston, Oleksii Kuchaiev, Ganesh
  Venkatesh, et~al.
\newblock Mixed precision training.
\newblock {\em arXiv preprint arXiv:1710.03740}, 2017.

\bibitem[\protect\citeauthoryear{{mlcommons}}{2023}]{A100}
{mlcommons}.
\newblock Inference: Datacenter v3.0 results.
\newblock \url{https://mlcommons.org/en/inference-datacenter-30/}, 2023.

\bibitem[\protect\citeauthoryear{MOBILE}{2019}]{Pytorch}
PYTORCH MOBILE.
\newblock End-to-end workflow from training to deployment for ios and android
  mobile devices.
\newblock \url{https://pytorch.org/mobile/home/}, 2019.

\bibitem[\protect\citeauthoryear{Patil \bgroup \em et al.\egroup }{2022}]{Poet}
Shishir~G Patil, Paras Jain, Prabal Dutta, Ion Stoica, and Joseph Gonzalez.
\newblock Poet: Training neural networks on tiny devices with integrated
  rematerialization and paging.
\newblock In {\em International Conference on Machine Learning}, pages
  17573--17583. PMLR, 2022.

\bibitem[\protect\citeauthoryear{Project}{2020}]{nnapi}
Android Open~Source Project.
\newblock Nnapi: Neural networks api.
\newblock {\em Android Developer Documentation}, 2020.

\bibitem[\protect\citeauthoryear{Sandler \bgroup \em et al.\egroup
  }{2018}]{mobilenetv2}
Mark Sandler, Andrew Howard, Menglong Zhu, Andrey Zhmoginov, and Liang-Chieh
  Chen.
\newblock Mobilenetv2: Inverted residuals and linear bottlenecks.
\newblock In {\em Proceedings of the IEEE conference on computer vision and
  pattern recognition}, pages 4510--4520, 2018.

\bibitem[\protect\citeauthoryear{Singapuram \bgroup \em et al.\egroup
  }{2022}]{Swan}
Sanjay Sri~Vallabh Singapuram, Fan Lai, Chuheng Hu, and Mosharaf Chowdhury.
\newblock Swan: A neural engine for efficient dnn training on smartphone socs.
\newblock {\em arXiv preprint arXiv:2206.04687}, 2022.

\bibitem[\protect\citeauthoryear{Sohoni \bgroup \em et al.\egroup
  }{2019}]{sohoni2019low}
Nimit~S Sohoni, Christopher~R Aberger, Megan Leszczynski, Jian Zhang, and
  Christopher R{\'e}.
\newblock Low-memory neural network training: A technical report.
\newblock {\em arXiv preprint arXiv:1904.10631}, 2019.

\bibitem[\protect\citeauthoryear{{Statista Inc.}}{2021}]{RAM}
{Statista Inc.}
\newblock Mobile ram usage worldwide from 1q'19 to 1q'21 (in gb per device).
\newblock
  \url{www.statista.com/statistics/1057679/mobile-ram-usage-worldwide-by-average-size-per-device/},
  2021.

\bibitem[\protect\citeauthoryear{Steiner \bgroup \em et al.\egroup
  }{2022}]{olla}
Benoit Steiner, Mostafa Elhoushi, Jacob Kahn, and James Hegarty.
\newblock Olla: Optimizing the lifetime and location of arrays to reduce the
  memory usage of neural networks.
\newblock {\em arXiv preprint arXiv:2210.12924}, 2022.

\bibitem[\protect\citeauthoryear{Sun \bgroup \em et al.\egroup
  }{2020}]{mobilebert}
Zhiqing Sun, Hongkun Yu, Xiaodan Song, Renjie Liu, Yiming Yang, and Denny Zhou.
\newblock Mobilebert: a compact task-agnostic bert for resource-limited
  devices.
\newblock {\em arXiv preprint arXiv:2004.02984}, 2020.

\bibitem[\protect\citeauthoryear{Tankard}{2016}]{policy_GDPR}
Colin Tankard.
\newblock What the gdpr means for businesses.
\newblock {\em Network Security}, 2016(6):5--8, 2016.

\bibitem[\protect\citeauthoryear{Tian \bgroup \em et al.\egroup
  }{2022}]{tian2022harmony}
Chunlin Tian, Li~Li, Zhan Shi, Jun Wang, and ChengZhong Xu.
\newblock Harmony: Heterogeneity-aware hierarchical management for federated
  learning system.
\newblock In {\em 2022 55th IEEE/ACM International Symposium on
  Microarchitecture (MICRO)}, pages 631--645. IEEE, 2022.

\bibitem[\protect\citeauthoryear{Wang \bgroup \em et al.\egroup }{2018}]{int8}
Naigang Wang, Jungwook Choi, Daniel Brand, Chia-Yu Chen, and Kailash
  Gopalakrishnan.
\newblock Training deep neural networks with 8-bit floating point numbers.
\newblock {\em Advances in neural information processing systems}, 31, 2018.

\bibitem[\protect\citeauthoryear{Wang \bgroup \em et al.\egroup }{2022}]{Melon}
Qipeng Wang, Mengwei Xu, Chao Jin, Xinran Dong, Jinliang Yuan, Xin Jin, Gang
  Huang, Yunxin Liu, and Xuanzhe Liu.
\newblock Melon: Breaking the memory wall for resource-efficient on-device
  machine learning.
\newblock In {\em Proceedings of the 20th Annual International Conference on
  Mobile Systems, Applications and Services}, pages 450--463, 2022.

\bibitem[\protect\citeauthoryear{Xu \bgroup \em et al.\egroup
  }{2022}]{Mandheling}
Daliang Xu, Mengwei Xu, Qipeng Wang, Shangguang Wang, Yun Ma, Kang Huang, Gang
  Huang, Xin Jin, and Xuanzhe Liu.
\newblock Mandheling: Mixed-precision on-device dnn training with dsp
  offloading.
\newblock In {\em Proceedings of the 28th Annual International Conference on
  Mobile Computing And Networking}, pages 214--227, 2022.

\bibitem[\protect\citeauthoryear{Zeng \bgroup \em et al.\egroup
  }{2020}]{AItask_4}
Xiao Zeng, Biyi Fang, Haichen Shen, and Mi~Zhang.
\newblock Distream: scaling live video analytics with workload-adaptive
  distributed edge intelligence.
\newblock In {\em Proceedings of the 18th Conference on Embedded Networked
  Sensor Systems}, pages 409--421, 2020.

\bibitem[\protect\citeauthoryear{Zhang \bgroup \em et al.\egroup
  }{2022}]{PilotFish}
Wei Zhang, Binghao Chen, Zhenhua Han, Quan Chen, Peng Cheng, Fan Yang, Ran Shu,
  Yuqing Yang, and Minyi Guo.
\newblock $\{$PilotFish$\}$: Harvesting free cycles of cloud gaming with deep
  learning training.
\newblock In {\em 2022 USENIX Annual Technical Conference (USENIX ATC 22)},
  pages 217--232, 2022.

\bibitem[\protect\citeauthoryear{Zhang \bgroup \em et al.\egroup
  }{2023}]{zhang2023memory}
Kai Zhang, Yutong Dai, Hongyi Wang, Eric Xing, Xun Chen, and Lichao Sun.
\newblock Memory-adaptive depth-wise heterogenous federated learning.
\newblock {\em arXiv preprint arXiv:2303.04887}, 2023.

\bibitem[\protect\citeauthoryear{Zhou \bgroup \em et al.\egroup }{2021}]{Octo}
Qihua Zhou, Song Guo, Zhihao Qu, Jingcai Guo, Zhenda Xu, Jiewei Zhang, Tao Guo,
  Boyuan Luo, and Jingren Zhou.
\newblock Octo: Int8 training with loss-aware compensation and backward
  quantization for tiny on-device learning.
\newblock In {\em USENIX Annual Technical Conference}, pages 177--191, 2021.

\end{thebibliography}
